\documentstyle[aps,pre,psfig,eqsecnum]{revtex}
\input epsf
\begin{document}

\draft
\twocolumn[\hsize\textwidth\columnwidth\hsize\csname@twocolumnfalse\endcsname

\title{Dynamic Entropy as a Measure of Caging and Persistent Particle
Motion in Supercooled Liquids}
\author{Paolo Allegrini, Jack F. Douglas and Sharon C. Glotzer$^*$}
\address{Polymers Division and Center for Theoretical and Computational
Material Science,\\ National Institute of Standards and Technology, 
Gaithersburg, MD, 20899, USA}
 
\date{Phys. Rev. E: submitted 12/18/98; revised 07/20/99}
\maketitle

\begin{abstract}
The length-scale dependence of the dynamic entropy is studied in a
molecular dynamics simulation of a binary Lennard-Jones liquid above
the mode-coupling critical temperature $T_c$.  A number of methods
exist for estimating the entropy of dynamical systems and we utilize
an approximation based on calculating the mean first-passage time
(MFPT) for particle displacement because of its tractability and its
accessibility in real and simulation measurements. The MFPT dynamic
entropy $S(\epsilon)$ is defined to equal the inverse of the average
first-passage time for a particle to exit a sphere of radius
$\epsilon$. This measure of the degree of chaotic motion allows us to
identify characteristic time and space scales and to quantify the
increasingly correlated particle motion and intermittency occurring in
supercooled liquids.  In particular, we identify a ``cage'' size
defining the scale at which the particles are transiently localized,
and we observe persistent particle motion at intermediate length
scales beyond the scale where caging occurs.  Furthermore, we find
that the dynamic entropy at the scale of one interparticle spacing
extrapolates to zero as the mode-coupling temperature $T_c$ is
approached.
\end{abstract}


\vskip2pc]
\narrowtext

\section{Introduction}

It has been suggested that the glass transition in cooled liquids is a
dynamic transition from an ergodic to a non-ergodic state. For
example, the ideal mode-coupling theory predicts that the molecules of
simple liquids become increasingly ``caged'' by surrounding molecules,
resulting in an ergodic to non-ergodic transition at a critical
temperature $T_c$ at which the fluid molecules become permanently
localized (i.e., the self-diffusion coefficient vanishes)
\cite{mc}. Although a tendency toward particle localization for
increasingly long times has been observed in simulations and
experiments on supercooled liquids, particle localization and
structural arrest does not actually occur at the extrapolated
temperature $T_c$ because the particles are eventually able to
``escape'' their cages.  Recent simulations have also shown the
tendency for particle motion to occur in an increasingly correlated
way in supercooled liquids \cite{ddkppg,dgpkp,kdppg,gd,gdp,dgp,bdbg},
a feature emphasized by the older, phenomenological Adam-Gibbs model
of glass formation \cite{ag65}.  The observed greater particle
mobility near $T_c$ is presumably a consequence of the increased
collective motion of cooled liquids (``hopping'' in the extended
version of the mode-coupling theory \cite{mc}) which restores the
ergodicity of the liquid for some temperature range below $T_c$. This
thermally activated collective motion apparently postpones the ergodic
to nonergodic transition to a lower temperature. In the Adam-Gibbs
model \cite{ag65}, this lower temperature corresponds to the
conjectured ``ideal'' glass transition temperature $T_o$, where the
equilibrium configurational entropy extrapolates to zero \cite{gdm}.

If glass formation indeed represents an ergodic to non-ergodic dynamic
transition, then it is important to define a dynamical measure of
order that quantifies both the ``closeness'' of the transition
\cite{tmk89}, and the degree of correlated motion in an equilibrium
glass-forming liquid.  Ergodic theory provides us with a natural
measure in the form of the ``dynamic entropy''
\cite{krylov,k59,GW,G2,Z81,L73}.

The concept of dynamic entropy was
introduced by Shannon in his theory describing the capacity of ideal
communication devices to transmit information \cite{shannon}. This
idea was later developed by Kolmogorov and others \cite{k59} into a general
measure of the ``degree of randomness'' or ``degree of chaos'' of
dynamical systems. According to Pessin's theorem \cite{pessin}, the
Kolmogorov-Sinai dynamic entropy $h_{KS}$ for a Hamiltonian dynamical
system equals the sum of the positive Lyapunov exponents \cite{pessin,livi}. 
These exponents are measures of the ``instability'' of the
system evolution \cite{krylov,Z81}. Dynamic entropy extends 
the equilibrium definition of entropy from statistical
mechanics to the {\it time domain}.
The dynamical entropy provides an estimate of the rate of growth of
``information'' (per unit time) required to describe the evolution of
a dynamical system \cite{GW,G2,D98} and it is also a measure of
the ``complexity'' of a dynamical system \cite{complexity}. The
dynamic entropy characteristically decreases as a system orders and
its exploration of its phase space becomes more restricted
\cite{cleary,butera}. Thus, the dynamic entropy decreases as a fluid
crystallizes or a spin system orders\cite{butera,posch,caiani,wales}.

The Kolmogorov-Sinai dynamic entropy has some shortcomings in the
description of complex configurational changes that occur in
supercooled liquids. In particular, $h_{KS}$ diverges for the ideal
process of Brownian motion (due to the non-differentiability of the
trajectories) \cite{GW,berger,wiener}. Consequently, we must anticipate
difficulties in applying dynamic entropy to quantify particle motions
at large length and time scales in the case of supercooled
liquids. Recently, there has been an important generalization of the
dynamic entropy concept that provides a ``bridge'' between
microscopic dynamical system descriptions and macroscopic stochastic
descriptions of liquid dynamics. This generalization recognizes that
the amount of information required to describe the paths of a
stochastic process depends strongly on the length scale of observation
$\epsilon$. The $\epsilon$-dependent dynamic entropy $h(\epsilon)$ of
Gaspard and Wang and others \cite{GW} (also called
``$\epsilon$-entropy'') reduces to the $h_{KS}$ entropy in the limit
of small $\epsilon$,
\begin{eqnarray} \label{liks}
\lim_{\epsilon \to 0} h(\epsilon) = h_{KS},
\end{eqnarray}
and is well-defined for idealized stochastic processes at a fixed,
nonvanishing $\epsilon$.  The dynamic entropy of a Brownian particle
$h_B(\epsilon)$ obeys the scaling relation
\begin{eqnarray} \label{hb}
h_B(\epsilon) \propto \epsilon^{-2},   \ \ \ \ \ \ \ \ \ \epsilon > 0,
\end{eqnarray}
where the proportionality constant is fixed by the particle diffusion
coefficient \cite{GW}. As mentioned above, $h(\epsilon)$ for
stochastic particle motion diverges as $\epsilon \to 0$ and the
exponent reflects the fractal dimension of the particle trajectories
\cite{GW,mandelbrodt}. Specifically, the exponent $2$ in Eq.~\ref{hb}
is the Hausdorff dimension of a Brownian path in three dimensions
\cite{hausdorff}, and in the limit of perfectly coherent (ballistic)
particle motion this exponent is $1$.  In idealized stochastic
processes ({\em e.g.} fractional Brownian motion, L\'evy flights,
etc.) the exponent in Eq.~1.2 can be identified with the path
Hausdorff dimension \cite{k59,GW,mandelbrodt,getoor}, and can take
values intermediate between 1 and 2.  This exponent reflects the
``degree of persistence'' in the particle displacement relative to
Brownian motion.

The scale dependent dynamic entropy $h(\epsilon)$ for complex
dynamical systems such as liquids depends strongly on the
observational scale $\epsilon$. At very small $\epsilon$ the
microscopic chaotic motion of the molecules is observed, so that
$h(\epsilon)$ varies slowly with $\epsilon$. The decorrelation of
particle velocities in a liquid occurs at a time and space scale
corresponding to the average interparticle ``collision time'' and
$h(\epsilon)$ starts varying with $\epsilon$ as this decorrelation
occurs. This helps us to identify a characteristic space and time
scale over which the bare microscopic dynamics can be coarse-grained
by a stochastic description. Correlations associated with particle
displacement arise at longer times in cooled liquids and $h(\epsilon)$
also helps us in determining the spatial and time scales over which
these correlations occur. $h(\epsilon)$ thus provides a measure of the
degree of chaotic motion appropriate to the description of real
systems at arbitrary observational scales and is an attractive
tool for quantifying the increasingly restricted motion in cooled
liquids.  It is notable that its definition is not restricted to
circumstances where statistical mechanical equilibrium exists, so that
this measure of the degree of chaos extends to non-equilibrium
situations such as the glass state and turbulent fluids \cite{GW}.

The calculation of $h(\epsilon)$ \cite{GW} (or $h_{KS}$) is generally
difficult, especially in cases where $h_{KS}$ is small and long
computational times are required for its accurate determination
\cite{GW,posch}. In the present paper, we utilize a simple
approximation for $h(\epsilon)$ that has the advantage of being
accessible in experiments on real materials and computer simulations
\cite{GW,gnature}. Provided that the spatial scale $\epsilon$ is not
too small \cite{GW}, $h(\epsilon)$ can be approximated by enclosing
the particle position at time $t=0$ by a sphere of radius $\epsilon$
centered on the particle, and then determining the time $\tau$ at
which the trajectory first arrives at the threshold distance
$\epsilon$ (see Fig.~1). We average this ``first-passage time'' over
all particles in the liquid to obtain the mean first-passage time
(MFPT) $\tau(\epsilon)$ and we define the ``MFPT dynamic entropy''
$S(\epsilon)$ as,
\begin{equation} \label{defentropy}
S(\epsilon) \equiv 1/\tau(\epsilon),
\end{equation}
where
\begin{equation} \label{deftau}
\tau(\epsilon) \equiv \int_{0}^{\infty}{dt}\, P_{\epsilon}(t) \, t.
\end{equation}
$P_{\epsilon}(t)dt$ is the probability that the particle reaches the
distance $\epsilon$ between $t$ and $t+dt$.  The dynamic entropy
$S(\epsilon)$ is thus one measure of the average ``escape rate'' of a
particle from its local environment \cite{jk}. We note that although
the definition of $S(\epsilon)$ is motivated by dynamical systems
theory concepts, this property defines an independently interesting
measure of correlated motion in liquids that does not rely on the
approximation relating $S(\epsilon)$ to $h(\epsilon)$.
\begin{figure}
\hbox to\hsize{\epsfxsize=0.65\hsize\hfil\epsfbox{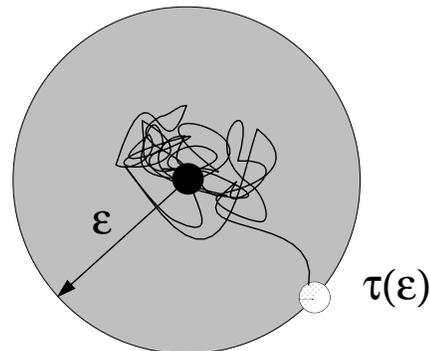}\hfil}
\caption{Schematic of a particle trajectory in a cooled liquid.  The
solid line represents the $\epsilon$-sphere (colored gray).
$\tau(\epsilon)$ is the first-passage time for the particle to reach
the sphere boundary.  Filled circle denotes initial particle position,
and open circle denotes particle at first-passage time.}
\label{fig0}
\end{figure}

In this paper we utilize $S(\epsilon)$ to identify characteristic
space and time scales in the particle dynamics and to quantify the
increasingly correlated motion observed in previous analyses of the
same simulations considered in the present paper
\cite{ddkppg,dgpkp}. These studies indicated the development of
large scale dynamical heterogeneity and the nature of this dynamical
heterogeneity has been examined in a series of recent papers that are
complementary to the present work
\cite{ddkppg,dgpkp,kdppg,gd,gdp,dgp}. There it was established that
transient clusters of highly ``mobile'' particles form in the cooled
liquid and that the average size of these clusters grows rapidly as
$T_c$ is approached \cite{dgpkp}. A pair distribution function for
particle displacements was defined and this quantity exhibits a
growing length scale upon cooling that reflects the clustering of
mobile particles \cite{gd,gdp,dgp}. The growing length scale is
time-dependent and attains a peak value at a time in the
$\alpha$-relaxation regime \cite{gdp,dgp}.

It has also been shown in the present liquid that the particles within
the mobile particle clusters move in cooperatively rearranging
``strings'' \cite{ddkppg,strings}.  Notably, the string-like
collective motion also begins well above $T_c$, but the strings
themselves exhibit no tendency to grow rapidly near $T_c$.  Instead,
the length distribution of the strings is found to be nearly
exponential and a similarity of this distribution to that commonly
observed in equilibrium polymerization has been noted
\cite{ddkppg}. Donati et al. \cite{ddkppg} have suggested that the
growing barrier height to particle motion is proportional to the
average string length, which would imply that these string-like
motions have a basic significance for understanding transport in
cooled liquids.  Thus, string-like correlated particle motion appears
to be an important mode of motion in our cooled liquid \cite{ddkppg}
and part of the motivation of the present work is to better
characterize the development of this type of collective motion. We are
also interested in the extent to which particle displacement becomes
intermittent in time in cooled liquids, since a growing intermittency
in particle motion has been suggested to underlie the glass transition
\cite{odagaki,goodguy}.

The paper is outlined as follows. In Section II we review some details
of the MD simulation data utilized in this work. Section III
examines $S(\epsilon)$ over a broad range of scales, and
dynamical regimes are defined where the motion is ballistic, transiently
localized, persistent and diffusive. These regimes are examined
in separate subsections.
We summarize our findings in Section IV.

\section{Simulation Details} 

The system studied is a three-dimensional binary mixture of 8000
Lennard-Jones (LJ) particles in which the sizes of the particles and
the interaction parameters are chosen to prevent crystallization and
demixing \cite{units}. The size of the A-particles is about 10\%
larger than that of the B-particles (while the mass is the same) and
the particles have a relative concentration 80:20 of A-particles to
B-particles. We report our results in dimensionless LJ units
\cite{units}.  The system was equilibrated at different temperatures
$T$ in the range (0.451,0.550). The density $\rho$ varied from $1.09$
particles per unit volume at the highest temperature to $1.19$ at the
lowest $T$ simulated.  For reference, the mode-coupling temperature
$T_c$ for this system is $T_c=0.435$ at $\rho \simeq 1.20$
\cite{ddkppg,dgp,kob}, so all the simulation data analyzed here is
well above the glass transition.  Configurational histories for up to
$4 \cdot 10^6$ molecular dynamics time-steps following equilibration
were stored for each run. Following equilibration in the NPT and NVT
ensembles, the trajectories were calculated in an NVE ensemble, and
snapshots containing the particle coordinates and velocities were
taken at logarithmic time intervals during the run. In this stage, the
equations of motion were integrated using the velocity Verlet
algorithm with a step size of 0.0015 at the highest temperature, and
0.003 at all other temperatures.  Adopting argon values for the LJ
parameters of the large particles implies an observation time of
$\approx 26$ ns for the coldest $T$. All data presented here are
calculated for the majority (A) particles only, except where otherwise
noted \cite{sametc}. Further details of the simulation can be found in
Ref.~\cite{dgpkp}.

\begin{figure}
\hbox to\hsize{\epsfxsize=1.0\hsize\hfil\epsfbox{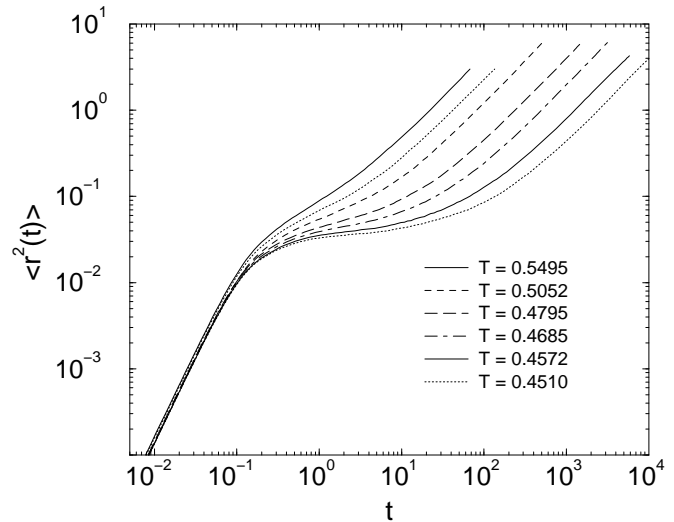}\hfil}
\caption{Mean square displacement of the majority species ($A$
particles) vs. time for different $T$.}
\label{figmsd}
\end{figure}

Over the temperature-density regime studied, the system exhibits the
usual features of a fragile \cite{angell}, glassforming liquid. For
example, the mean square displacement $\langle r^2(t) \rangle \equiv
\bigl \langle \frac{1}{N_A} \sum_{i=1}^{N_A} |{\bf r}_i(t) - {\bf
r}_i(0)|^2 \bigr \rangle$ for the A particles is shown in
Fig.~\ref{figmsd} for different T. Here ${\bf r}_i(t)$ is the position
of particle $i$ at time $t$, $N_A$ is the number of A particles
(6400), and $\langle \cdots \rangle$ denotes an ensemble average.  For
each state point, a ``plateau'' exists in both the mean square
displacement and the self-part of the intermediate scattering function
$F_s({\bf q},t)$ as a function of $t$ (see \cite{dgpkp}).  The plateau
in Fig.~\ref{figmsd} separates an early time ``ballistic'' regime from
a late time diffusive regime. The plateau is interpreted as implying
``caging'' of the particles and this phenomenon is typical for liquids
of low temperature or high density.  Over the range of $T$ studied,
the $\alpha$-relaxation time $\tau_{\alpha}$, describing the decay of
$F_s({\bf q},t)$ (at the value of $q$ corresponding to the first peak
in the static structure factor), increases by 2.4 orders of magnitude,
and follows a power law $\tau_{\alpha} \sim (T-T_c)^{-\gamma}$, with
$T_c \simeq 0.435$ and $\gamma \simeq 2.8$. The simulated
liquid states analyzed here therefore exhibit relaxation behavior
characteristic of a supercooled liquid.  No long range structural
correlations due to density or composition fluctuations are apparent
in the simulation data \cite{dgpkp}.

\section{Characteristic decorrelation time and space scales}

\begin{figure}
\hbox to\hsize{\epsfxsize=1.0\hsize\hfil\epsfbox{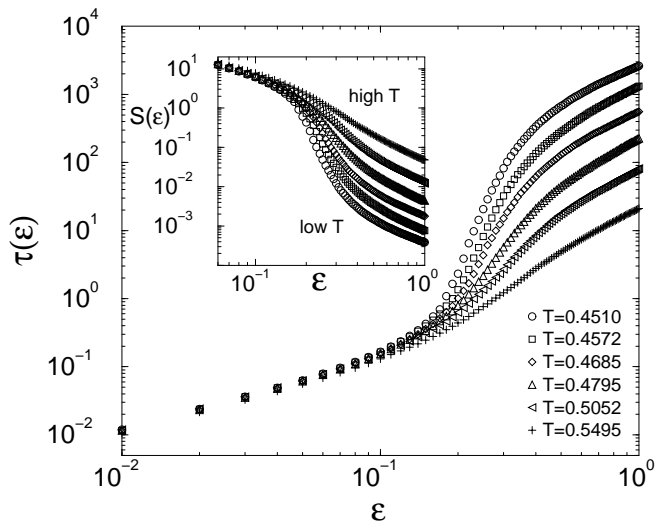}\hfil}
\caption{Mean first-passage time $\tau$ of the majority species (A
particles) versus $\epsilon$, for different $T$. Inset: Mean
first-passage time dynamic entropy $S(\epsilon)$. Compare the
$S(\epsilon)$ variation observed here for a cooled liquid with the
dynamic entropy $h(\epsilon)$ calculated for a one-dimensional model
map exhibiting diffusion at long times (see Fig.~25b of Gaspard and
Wang [8]).}
\label{figtaueps}
\end{figure}

In Fig.~\ref{figtaueps} we show the MFPT $\tau(\epsilon)$ plot for six
different runs corresponding to varying the temperature of the system
from $T=0.550$ to $T=0.451$.  In the inset we show the dynamic entropy
$S(\epsilon) \equiv 1/\tau(\epsilon)$. Note that the variation of
$\tau(\epsilon)$ with $\epsilon$ exhibits similar qualitative trends
to the variation of $t$ with $\langle r^2(t)\rangle$ shown in
Fig.~\ref{figmsd}.

For small $\epsilon$, corresponding to the inertial regime,
$S(\epsilon)$ is insensitive to temperature. At intermediate
$\epsilon$ values we see a decrease of $S(\epsilon)$ with decreasing
$T$ and an increase in the magnitude of the slope in the log-log plot.
A strong temperature dependence of $S(\epsilon)$ is apparent at a
scale on the order of one interparticle distance ($\epsilon=1$). On
these larger scales, we show in a later subsection that $S(\epsilon)$
exhibits a power-law scaling with $\epsilon$ and reduced temperature.

\begin{figure}
\hbox to\hsize{\epsfxsize=1.0\hsize\hfil\epsfbox{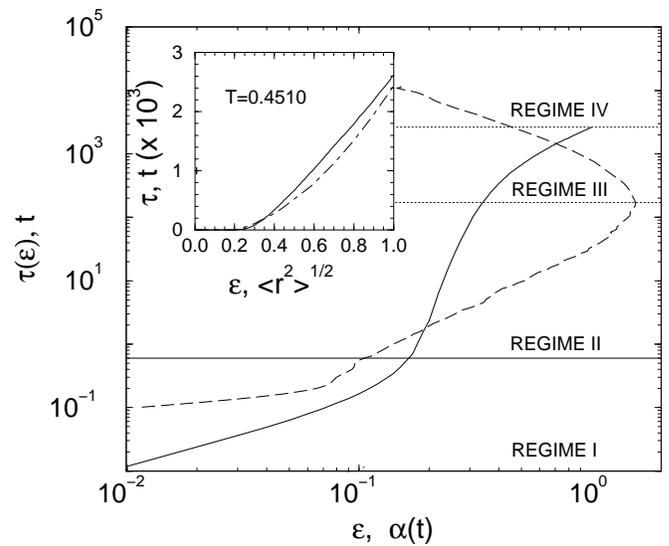}\hfil}
\caption{Classification of dynamic regimes. MFPT $\tau(\epsilon)$
(solid curve) plotted versus $\epsilon$, and the non-Gaussian
parameter $\alpha(t)$ (dashed curve) plotted versus t, for $T=0.451$.
Inset: Comparison between $\tau(\epsilon)$ (solid curve) and $\langle
r^2(t)\rangle$ (dot-dashed curve) on a linear scale.}
\label{figmfpt}
\end{figure}

It is apparent from Fig.~\ref{figmsd} that the particle displacement
in this cooled liquid is not Brownian over most of the simulation time
scales, and it is conventional to quantify this deviation by a
``non-Gaussian parameter'' $\alpha (t)$ involving the moments $\langle
r^2 (t) \rangle$ and $\langle r^4 (t) \rangle$ of the self part of the
van Hove correlation function $G_s(r,t) \equiv \Bigl \langle
\frac{1}{N_A}\sum_{i=1}^{N_A} \delta \bigl( {\bf r} - ({\bf r}_i(t) -
{\bf r}_i(0)) \bigr) \Bigr \rangle.$ The parameter $\alpha(t)$ is
defined as,
\begin{equation}
\alpha(t)
\equiv \frac{3 \langle r^4(t) \rangle}{5\langle r^2(t) \rangle^2} -1,
\end{equation}
and vanishes for Brownian motion.  $\alpha(t)$ is shown in
Fig.~\ref{figmfpt} together with $\tau(\epsilon)$ for the coldest run
($T=0.451$); note that for $\alpha(t)$, time $t$ is plotted on the
ordinate axis.  This comparison allows us to identify four regimes: An 
inertial regime (Regime I) where the non-Gaussian parameter is small;
a ``localization'' regime characterized by a large value for the slope
in the $\tau(\epsilon)$ log-log plot and by a growing $\alpha(t)$
(Regime II); a regime of particle motion that is persistent relative
to Brownian motion, and where $\alpha(t)$ decreases (Regime III); and
a fourth regime where the non-Gaussian parameter has decayed back to
very small values so that the particle motion is nearly Brownian
(Regime IV).  The four regimes are examinated in detail in the
following subsections.

We emphasize that while some parallelism exists between the
mean square displacement and the first-passage time, the inset of
Fig.~\ref{figmfpt} comparing these quantities shows that they are not
equivalent.  However, if the distribution of particle displacments
$G_s(r,t)$ were always exactly Gaussian (i.e. $\alpha(t)=0$ for every
$t$), then a simple inverse function relation should hold between
these quantities.  $G_s(r,t)$ obeys a ``scaling relation'' if we can
rescale $G_s(r,t)$ as,
\begin{equation} \label{diffscaling}
G_s(r,t)=\frac{1}{t^{\nu}}f \left( \frac{r}{t^{\nu}} \right),
\end{equation} 
where $f$ is some function.  Simple dimensional analysis based on
(\ref{diffscaling}) implies,
\begin{equation}
\langle r^2 (t) \rangle \propto t^{2\nu}
\end{equation}
and the scaling of the first-passage time with $\epsilon$,
\begin{equation}
\tau(\epsilon) \propto \epsilon^{1/\nu}.
\end{equation}

In practice, these idealized scaling relations are restricted to
certain time and space scales.  Scaling with $\nu=1$ is observed in the
short time inertial regime.  This result can be inferred from the
general relation between the second moment of $G_s(r,t)$ and the
particle velocity,
\begin{equation} \label{phix2}
\langle {\dot {\bf r}} (0) \cdot {\dot {\bf r}}(t) \rangle 
= \frac{1}{2}\frac{d^2 \langle r^2 (t) \rangle}{dt^2}, 
\end{equation}
and the constancy of the velocity autocorrelation function at short
times \cite{lee,equipartition}, 
\begin{equation} 
\langle {\dot {\bf r}} (0) \cdot {\dot {\bf r}}(t) \rangle 
\to \langle {\dot {\bf r}} (0) \cdot {\dot {\bf r}}(0) \rangle
\equiv v_o^2 \ \ \ \ \ t \to 0, 
\end{equation}
where $v_o$ is the average particle velocity. Integrating
Eq.~\ref{phix2} over a short time interval obviously gives $\langle
r^2(t) \rangle \sim t^2$ or ``ballistic-like'' motion. Actually, this
scaling is just a consequence of the existence of equilibrium, and
should not be construed as necessarily implying the absence of
interparticle interactions at short timescales \cite{equipartition}. A
Gaussian form for the van Hove correlation function in this short time
regime is ensured by the Maxwell-Boltzmann distribution for the
particle velocities. In the opposite extreme of very long times, the
central limit theorem governing the sum of independent particle
displacements implies that the particle displacement distribution is
Gaussian, and that $\nu=1/2$. Transient scaling regimes can be
observed at intermediate time scales, however. We refer to particle
displacements as ``persistent'' relative to Brownian motion if $\nu >
1/2$, or ``localized'' relative to Brownian motion if $\nu < 1/2$.

\subsection{Inertial regime}
In the limit of very small $\epsilon$ we probe the fast microscopic
dynamics associated with the decorrelation of the particle momenta.
It is difficult to probe this decorrelation directly using the
first-passage time approximation to the dynamic entropy.
Fig.~\ref{figtaueps} indicates that our approximation for the dynamic
entropy appears to diverge for $\epsilon \rightarrow 0$, so our
approximation must break down in this limit. Gaspard and Wang have
pointed out before that the first-passage time approximation to the
dynamic entropy breaks down in this limit \cite{GW}, so this
shortcoming is to be expected.  An estimate of the expected plateau in
$S(\epsilon)$ corresponding to the Kolmogorov-Sinai entropy can be
obtained by determining a cut-off time of the first-passage time
distributions in the fast dynamics regime.

\begin{figure}
\hbox to\hsize{\epsfxsize=1.0\hsize\hfil\epsfbox{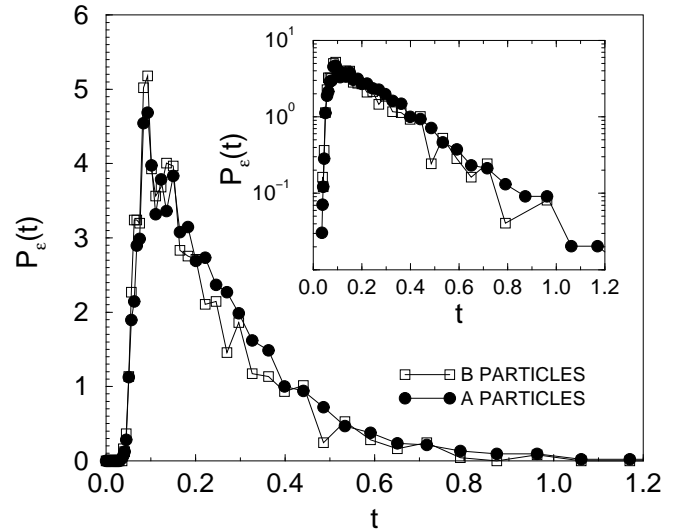}\hfil}
\caption{Probability of first-passage time $P_{\epsilon}(t)$ for the A
and B particles at $T=0.451$ and $\epsilon=0.1$. INSET: Same data 
represented in a semi-log scale.}
\label{figAeB}
\end{figure}

The dynamics in this regime is examined by setting the magnitude of
the first-passage sphere ($\epsilon$-sphere) about the center of each
particle to be small enough that a collision does not usually occur
before the particle leaves the $\epsilon$-sphere (see Fig.~1).  By
focusing on the particle first-passage time in this regime we identify
a time and space scale over which particle velocities begin
decorrelating. The first-passage time in this regime is insensitive to
the type (A or B) of particle since they have the same mass. This is
apparent in Fig.~\ref{figAeB} where the first-passage time
distributions for both the $A$ and $B$ particles are shown for the
coldest run and $\epsilon=0.1$.

\begin{figure}
\hbox to\hsize{\epsfxsize=1.0\hsize\hfil\epsfbox{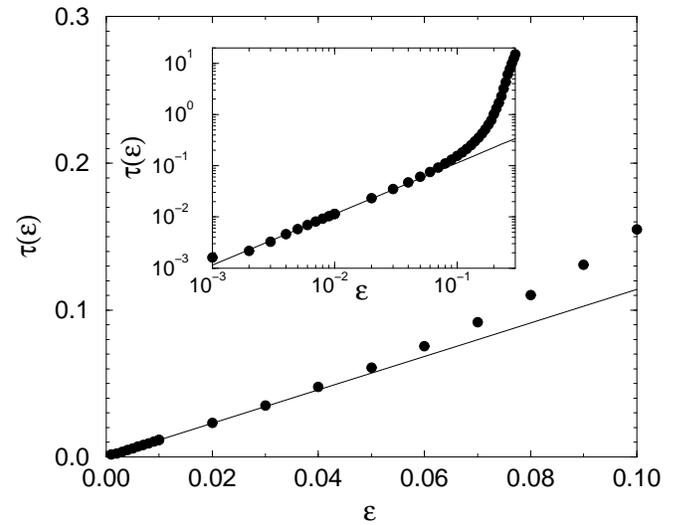}\hfil}
\caption{$\tau$ versus $\epsilon$ in the small $\epsilon$ regime,
$0.001 < \epsilon < 0.15$, for T=0.468.  Inset: log-log plot of the
same data. Note that the asymptotic linear scaling of
$\tau(\epsilon)=1.14\epsilon$ (solid lines) breaks down around a value
$\epsilon_v \approx 0.05$, corresponding to $\tau_v \approx 0.6$.
$\epsilon_v$ and $\tau_v$ are approximately independent of
temperature (see also Fig.~3).}
\label{smalleps}
\end{figure}

The idealization of ballistic particle motion implies the time
$\tau(\epsilon)$ it takes for the particle to exit the sphere of
radius $\epsilon$ equals $\epsilon/v_0$.  Fig \ref{smalleps} shows an
expanded plot of $\tau$ versus $\epsilon$ in the small $\epsilon$
regime where this linear dependence of $\tau$ on $\epsilon$ is
apparent. A non-linear dependence of $\tau$ on $\epsilon$ develops as
the particle velocities decorrelate at a ``velocity decorrelation
scale'' $\epsilon_v \approx 0.05$. This distance corresponds to the
time $\tau_v \approx 0.6$ (on the order of $10^{-13}s$ in argon
units). We find that $\epsilon_v$ and $\tau_v$ are approximately
independent of temperature in the present simulation (see also
Fig.~3).

\begin{figure}
\hbox to\hsize{\epsfxsize=1.0\hsize\hfil\epsfbox{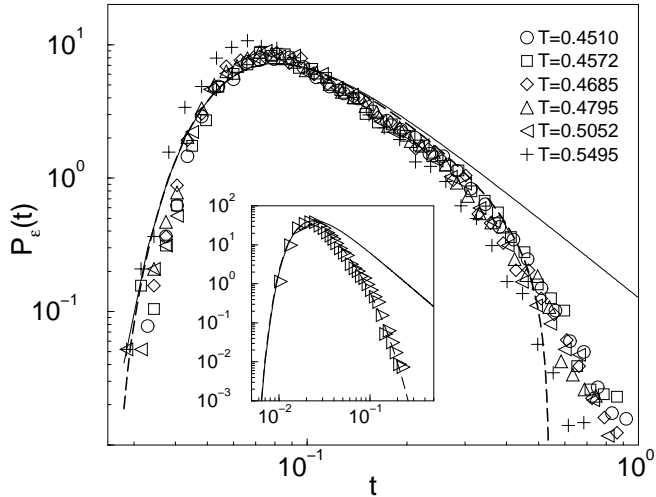}\hfil}
\label{cutoff}
\caption{First-passage time distributions $P_{\epsilon}(t)$ in the
inertial regime. Main figure contains six different temperatures at
$\epsilon=0.1$ plotted log-log. Solid line indicates Eq.~3.7 where
the constant of proportionality has been adjusted to best fit the
data. The long-dashed line indicates the cut-off distribution function
Eq.~3.8 with no free parameters.  Inset: $P_{\epsilon}(t)$ at $T=
0.468$ for $\epsilon=0.03$. This value of $\epsilon$ is less than the
velocity decorrelation scale $\epsilon_v$ indicated in
Fig.~6. Eqs.~\ref{wsmalle} and \ref{rsmalle} are also shown in
comparison with the simulation data.  Note the absence of the long
tail in the inset distribution.}
\end{figure}

We obtain further insight into the decorrelation of particle
velocities in the inertial regime by examining the distribution of
first-passage times, $P_{\epsilon}(t)$, as a function of
$\epsilon$. Fig.~7 shows $P_{\epsilon}(t)$ for $\epsilon=0.1$ and
$\epsilon=0.03$ for $T=0.468$.  We obtain a good approximation to
$P_{\epsilon}(t)$ in the inertial regime through a Gaussian
approximation for $G_s(r,t)$ \cite{mcquarrie} in conjunction with the
first-passage distribution for Brownian paths \cite{ciesielski}.  The
first-passage time distribution $P_{\epsilon}(t)$ scales as
\begin{equation} \label{wsmalle}
P_{\epsilon}(t) \sim 
\exp{\left({ \frac{-\epsilon^2}{2\langle R^2 \rangle}}\right)}
\end{equation} 
at short times and decays exponentially at long times,
$P_{\epsilon}(t) \sim \exp \left[ -t/\tau(\epsilon)\right]$
\cite{ciesielski}. We then introduce the approximation
\begin{equation}\label{rsmalle}
P_{\epsilon} \simeq \frac{A}{t^2} \exp \left[ -\frac{1}{2}
\left( \frac{\epsilon}{v_0 t} \right)^2
-\frac{t}{\tau(\epsilon)} \right],
\end{equation}
where $A$ is a normalization constant. $\langle R^2 \rangle^{1/2}$ in
Eq.~\ref{wsmalle} has been replaced by $v_ot$ based on the assumption
that the particle displacements are nearly ``ballistic'' in the
inertial regime.  Note that the limiting expression for
$P_{\epsilon}(t)$ given by Eq.~\ref{wsmalle} can also be deduced by
assuming a Maxwell velocity distribution (one-dimensional due to the
near rectilinear motion) with the velocity replaced by $\epsilon / t$.
Fig.~7 shows that Eq.~\ref{wsmalle} (solid line) does not provide an
accurate fit to the data using $v_0 = 1/1.14$ from the linear slope in
Fig.~\ref{smalleps}, while the approximate expression
Eq.~\ref{rsmalle} fits well for $\epsilon$ less than the ``velocity
decorrelation scale'' $\epsilon_v \approx 0.05$.  For $\epsilon >
\epsilon_v$, $P_{\epsilon}(t)$ shows evidence of developing a
power-law ``tail'' at long times, as seen in the main part of Fig.~7.
This tendency becomes more developed at larger $\epsilon$, as
discussed in the next subsection.  The inset of Fig.~7 shows
first-passage time data for $\epsilon < \epsilon_v$ where the tail is
nearly absent and the cut-off is evident. At this point, we note that
in liquids $h_{KS}$ characteristically scales as the inverse of the
average interparticle collision time and thus has the interpretation
of a microscopic ``collision rate'' \cite {krylov,GW}. The time
$\tau_v$ can be considered as an average particle collision time so
that an inverse relation betwen $h_{KS}$ and $\tau_v$ is expected.  We
do not consider the relation between $\tau_v$ and $h_{KS}$, since
simulations over a broader temperature range and the direct
calculation of the Kolmogorov-Sinai entropy through Lyapunov exponent
spectra \cite{dzugutov,sri} are required for such an investigation.

\subsection{Particle localization regime}

The tendency of particle motion to become increasingly localized, as
emphasized by the mode-coupling theory, is a conspicuous feature of
experimental and simulation data on supercooled liquids. The
``plateau'' in the mean square displacement log-log plots (see Fig.~2)
indicates transient particle localization or ``caging'', and the
persistence of this plateau increases with decreasing $T$
\cite{doliwa}. Next, we utilize the dynamic entropy concept to
quantify this particle localization.

\begin{figure}
\hbox to\hsize{\epsfxsize=1.0\hsize\hfil\epsfbox{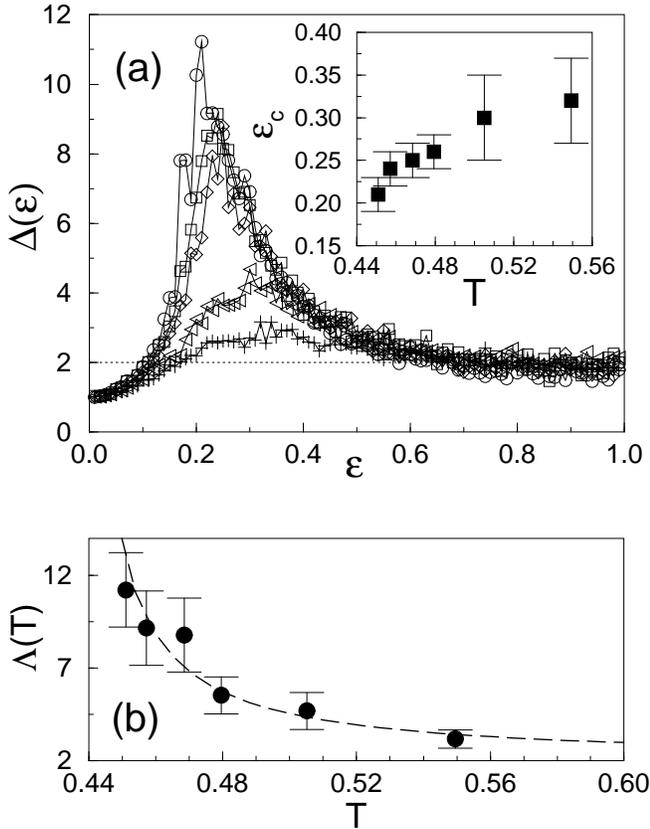}\hfil}
\caption{(a) $\Delta(\epsilon) \equiv d \log \tau(\epsilon)/d \log
\epsilon)$ for (from top to bottom) $T=0.451, 0.457, 0.468, 0.480,
0.505$, and $0.550$. Inset: ``Cage'' size $\epsilon_c$ versus $T$. (b)
$\Lambda \equiv \max \Delta(\epsilon)$ plotted versus $T$.  The
dashed line is a power law fit to the data as indicated in the text.}
\label{logder}
\end{figure}

An increase in the slope of $\log{\tau(\epsilon)}$ versus
$\log{\epsilon}$ in Fig.~\ref{figtaueps} provides evidence for
localization. A numerical differentiation of the data in
Fig.~\ref{figtaueps} is shown in Fig.~\ref{logder}a where
$\Delta(\epsilon)$ denotes the logarithmic derivative
$\Delta(\epsilon) \equiv d \log \tau/d \log \epsilon$.  The scaling
behavior in Eq.~3.4 implies that $\Delta(\epsilon)$ corresponds to the
fractal dimension $1/\nu$ of the particle trajectories. Fig.~8a shows
that for all $T$, $\Delta(\epsilon) \to 1$ for small $\epsilon$, and
$\Delta(\epsilon) \to 1/\nu$ for large $\epsilon$.  At intermediate
values of $\epsilon$, we observe that $\Delta(\epsilon)$ develops a
maximum at $\epsilon_c$, corresponding to the inflection point in
Fig.~\ref{figtaueps}. This length scale defines a distance that is
difficult for the particle to exceed, and thus we define $\epsilon_c$
as the ``cage'' size. The inset in Fig.~\ref{logder}a shows that
$\epsilon_c$ decreases with temperature, so that increased particle
confinement occurs with cooling. Independent evidence indicating the
significance of this characteristic scale is discussed later in this
subsection.

We denote the value of $\Delta(\epsilon)$ at $\epsilon_c$ by the
``localization parameter'' $\Lambda(T) \equiv \max \Delta(\epsilon) =
\Delta(\epsilon_c)$, and its $T$-dependence is shown in Fig
\ref{logder}b. The value of $\Lambda(T)$ increases with cooling,
consistent with increasing particle localization ($\nu < 1/2$).  The
relatively noisy data in Fig.~\ref{logder}b can be fitted by a power
law, $\Lambda(T) = 2+0.15(T-T_c)^{-1.03}$ with $T_c=0.435$.

\begin{figure}
\hbox to\hsize{\epsfxsize=1.0\hsize\hfil\epsfbox{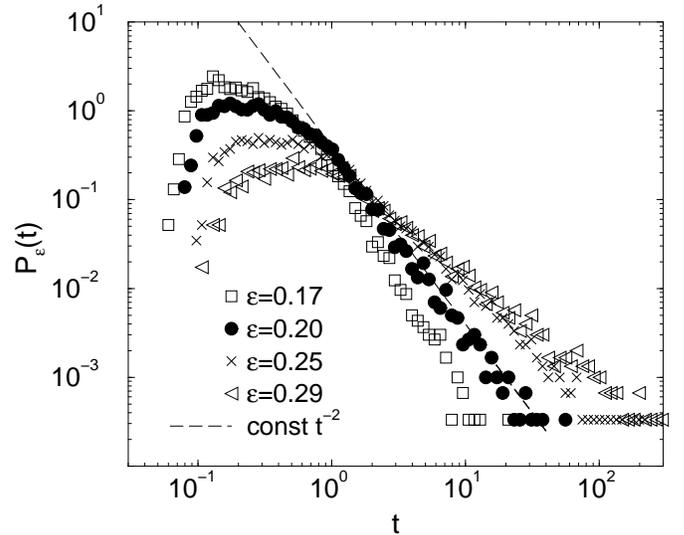}\hfil}
\caption{First-passage time distributions for $T=0.451$, at different
values of $\epsilon$ near $\epsilon_c = 0.21 \pm 0.02$.  The dashed line 
indicates an inverse power law $t^{-2}$ for comparison.}
\label{figdcsj}
\end{figure}

Our identification of the cage size $\epsilon_c$ from the maximum
value of $\Delta(\epsilon)$ is further supported by the examination of
the first-passage time distribution $P_{\epsilon}(t)$ for $\epsilon$
near $\epsilon_c$. We find that $P_{\epsilon}(t)$ develops a long time
power-law tail in this intermediate regime which is symptomatic of the
development of intermittency in particle motion and particle
localization \cite{odagaki}. In Fig.~\ref{figdcsj} we show for
$T=0.451$, $P_{\epsilon}(t)$ at several values of $\epsilon$ near
$\epsilon_c =0.21 \pm 0.02$.  The apparent power law for the
$P_{\epsilon}(t)$ tail varies with $\epsilon$ and has a value near 2
for $\epsilon \approx \epsilon_c$.

This power-law tail behavior in $P_{\epsilon_c}(t)$ is shared by the
four coldest temperatures, as shown in Fig.~10, where the asymptotic
behavior of the distributions is seen to be numerically very
similar. The difference in their first moments
[i.e. $\tau(\epsilon_c)$ in Fig.~3] reflects both the differences
between the distributions in Fig.~10 at short times, and the
asymptotic cut-off in the distributions that is difficult to resolve
numerically.

\begin{figure}
\hbox to\hsize{\epsfxsize=1.0\hsize\hfil\epsfbox{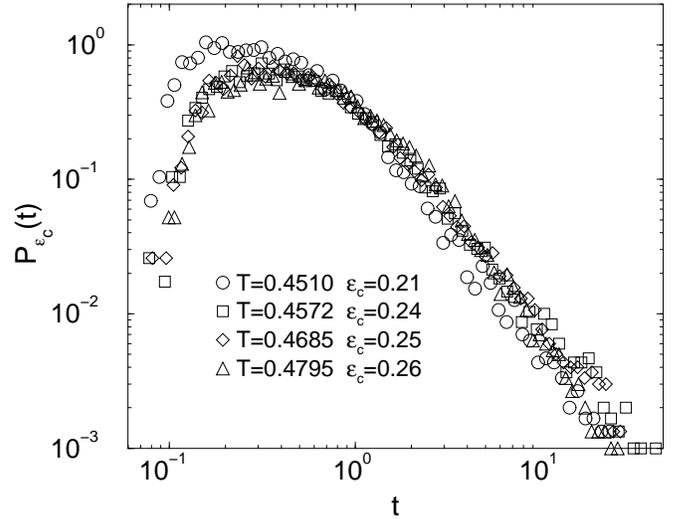}\hfil}
\caption{First-passage time distributions at the cage scale,
$\epsilon=\epsilon_c$, for four different $T$.}
\label{figepsc}
\end{figure}

Odagaki \cite{odagaki} has suggested that the second and first moments
of the first-passage time distribution at the scale of one
interparticle distance diverge at $T_c$ and at the glass transition
temperature $T_g$, respectively, and a recent paper by Hiwatari and
Muranaka \cite{hiwatari} supports this glass transition scenario.  A
transition to intermittent particle motion at the glass transition has
also been suggested by Douglas and Hubbard \cite{goodguy}.  Our
observations are consistent with the growth of intermittency of
particle motion as $T_c$ is approached, but we cannot confirm
theoretical predictions of a dynamical transition in the degree of
intermittancy until lower temperatures are examined.  An accurate test
of these predictions will require carefully equilibrated data below
$T_c$, beyond the temperature range of the simulations analyzed here.
We point out that, in the present system, the scale at which this
intermittency occurs is substantially {\it smaller} than one
interparticle spacing, and instead pertains to motion at the scale of
the cage size, $\epsilon_c$.

\subsection{Regime of persistent particle motion}
Previous work has shown that particle motion becomes increasingly
collective in this supercooled liquid and that an important mode of
motion at the scale of the interparticle distance involves the
string-like collective motion of particles. A visualization of this
process \cite{movie} suggested to us that this process
becomes increasingly ``coherent'' or ``jump-like'' at lower
temperatures and this tendency towards ``coherent jumping''
has been noticed in a number of other physical systems (melting
of hard disks \cite{alder}, hexatic liquids \cite{murray} and ordering
plasmas \cite{choquard}). In this subsection we utilize $S(\epsilon)$ 
to further quantify this effect.

Regimes I and II were defined by characteristic spatial scales at
which changes occur in the first-passage time distributions.  However,
the long run times of the simulations analyzed here necessitated the
storing of configurations on a logarithmic, rather than linear, time
scale, for all but the coldest simulation \cite{dgpkp}. As a
consequence, first-passage time distributions cannot be obtained over
a continuous range of $\epsilon$ for large $\epsilon$.  In the absence
of this information, we roughly identify Regime III by the tendency
for the non-Gaussian parameter $\alpha(t)$ to decrease (see
Fig.~4). In Fig.~11 we show the apparent fractal dimension
$\Delta(\epsilon)$ in Regime III for the highest and lowest
temperatures. (The results for all $T$ are shown over an extended
scale in Fig.~8).  Notice that persistent particle motion
($\Delta(\epsilon)<2$) develops for $\epsilon > 0.6$ in Fig.~8
(although $\Delta$ remains near 2 at the highest temperatures) and
consequently we show $\Delta(\epsilon)$ for the range $0.7 < \epsilon
< 1.0$ in Fig.~11.  The value of $\epsilon$ at which
$\Delta(\epsilon)=2$ provides a more precise estimate for the
beginning of Regime III.  Although the data is noisy, we find that
$\Delta(\epsilon) \approx 2$ within numerical error, and is nearly
independent of $\epsilon$ for high $T$. Particle displacement at high
temperature is then reasonably approximated by Brownian motion on
these spatial scales.  However, a substantially smaller average value
of $\Delta(\epsilon) \approx 1.7$ (dashed line in Fig.~11) is found
for the lowest $T$.  Thus, we find the particle motion becomes
increasingly persistent on cooling.

Why is persistent motion not observed in the mean square displacement?
The tendency for persistent particle motion is not apparent in
$\langle r^2(t)\rangle$ shown in Fig.~\ref{figmsd} or in the inset of
Fig.~\ref{figmfpt} because, when averaging over the squared
displacements, the contribution of the few particles that are at any
given time moving persistently is ``washed out''.  However, these
particles give a large contribution to the mean first-passage time, so
that this quantity is therefore a sensitive indictor of persistent
particle motion.

\begin{figure}
\hbox to\hsize{\epsfxsize=1.0\hsize\hfil\epsfbox{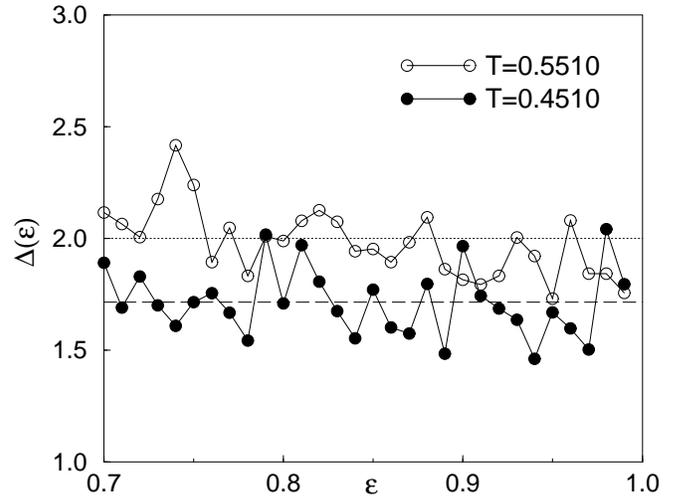}\hfil}
\caption{Apparent fractal dimension $\Delta(\epsilon)$ of
particle displacements in Regime III. From Fig.~8, highest and lowest
simulation temperature $T=0.5510$ and $T=0.4510$ are shown for $0.7 <
\epsilon < 1.0$.}
\label{figppm}
\end{figure}

We can obtain insight into the emergence of persistent particle motion
in our cooled liquid from idealized models of Brownian motion subject
to potential fluctuations.  If the potential fluctuations are
quenched, there is a tendency towards particle localization
\cite{grassberger}, but the occurrence of fluctuations in the
potential in both space and time can lead to persistent particle
motion.  In particular, if the fluctuations are delta-correlated in
both space and time, then the exponent $\Delta$ equals $3/2$
\cite{kardar}.  The effects observed in Ref.~\cite{kardar} are
qualitatively consistent with our understanding of the origin of
correlated motion.  At short times, the existence of relatively
immobile particles leads to a randomly fluctuating field felt by those
particles free to move at a given point in time.  This spatially
fluctuating field is responsible for the particle caging or
localization on timescales short compared to the decorrelation time
$\tau_{\alpha}$ of the ``structural fluctuations'' (associated with
the relatively immobile particles).  At longer times, the formerly
immobile particles become mobile and the potential field fluctuates in
time, leading to an enhancement in the particle displacement.  At
still longer times, thermal fluctuations restore equilibrium and
particle displacement ultimately becomes diffusive.

Finally, we point out that particle motion can be persistent even in
the absence of a secondary (so-called ``hopping'') peak in $G_s(r,t)$.
Instead, persistent particle motion contributes to a long tail in
$G_s(r,t)$ in the temperature range of the present system
\cite{dgpkp,kdppg}.  This long tail sharpens up and becomes a
secondary peak at lower $T$ \cite{secondpeak}, indicating the
increased contribution of collective particle motion to transport
below $T_c$.

\subsection{Large scale particle displacement}

Particle displacement in a liquid at large scales is described by
Brownian motion so that $S(\epsilon)$ should scale asymptotically as
$\epsilon^2$ for large $\epsilon$, regardless of temperature. It is
apparent in Fig.~\ref{figmsd} that the data in the asymptotic
diffusive regime is limited, especially at lower
temperatures. Previous work has shown that there is a tendency for 
``mobile'' particles, which dominate transport in cooled liquids near
$T_c$, to move an interparticle distance during the time in which they
are mobile \cite{ddkppg,dgpkp}. This happens because the particles
tend to move between local minima in the potential surface describing
the interparticle interaction \cite{minima,tbs}. This feature is
especially apparent in the string-like particle motion noted before,
where it has been observed that the strings ``disintegrate'' once an
interparticle displacement has been achieved \cite{ddkppg}. Thus, one
interparticle distance is taken to be the minimal scale of the large
scale particle displacement regime (Regime IV).

\begin{figure}
\hbox to\hsize{\epsfxsize=1.0\hsize\hfil\epsfbox{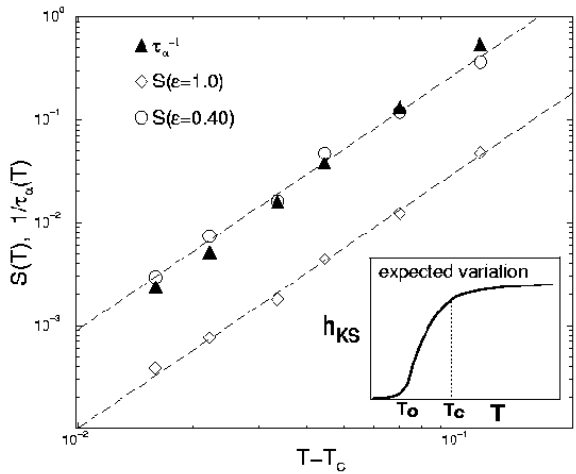}\hfil}
\caption{Dynamic entropy at the scale of one interparticle separation,
$S(\epsilon=1)$, versus $T-T_c$, with $T_c=0.435$. Diamonds refer to
the simulation data for $S(\epsilon = 1)$, and the dashed lines refer
to the power law, $S(T) \sim (T-T_c)^{2.5}$. We also observe this
scaling for $S(\epsilon=0.4)$ (circles). The inverse of the structural
relaxation $1/\tau_{\alpha}$ (triangles) follows a power law
$1/\tau_{\alpha} \sim (T-T_c)^{2.8}$.  Inset: Schematic indication of
expected temperature variation of dynamic entropy $h_{KS}$ at small
scales. Note that the extrapolated ergodic-nonergodic transition in
the $\epsilon \to 0$ limit should occur at a temperature $T_o \ < \
T_c$.}
\label{figalpha1}
\end{figure}

In the previous discussion, we have established that $S(\epsilon)$ is
insensitive to temperature for small $\epsilon$ over the temperature
range investigated. This insensitivity accords with the expected
variation of dynamic entropy $h_{KS}$.  At higher $T$ we expect the
dynamic entropy to saturate, while a decrease should accompany the
more restricted particle motion at lower $T$ \cite{posch}. A variation
similar to that shown in the inset of Fig.~12 has been established for
the ordering of the XY model \cite{butera}.  The investigation of the
anticipated ergodic-nonergodic transition and its possible relation to
a vanishing of $h_{KS}$ requires the calculation of the full Lyapunov
spectrum, which is currently prohibitive for a system of the present
size.  However, we can investigate $S(\epsilon)$ at a larger scale on
the order of an interparticle spacing.  This should be interesting
because of the strong $T$-dependence in $S(\epsilon)$ at this length
scale noted above (see Fig.~3).

In Fig.~\ref{figalpha1} we examine the $T$-dependence of
$S(T;\epsilon)$ at the scale of one interparticle separation
$\epsilon=1$. We observe that $S(T;\epsilon=1)$ obeys a power law to a
good approximation over the temperature range investigated, and that
$S(T;\epsilon=1)$ seems to extrapolate to zero at the mode-coupling
temperature $T_c=0.435$.  As shown in Fig.~12, a reasonable fit to the
data is obtained with the relation
\begin{equation} \label{spowerlaw}
S(\epsilon=1) \sim (T-T_c)^{2.5}, \,\,\, T_c=0.435. 
\end{equation}
The scaling of $S(\tau;\epsilon=1)$ is compared in
Fig.~\ref{figalpha1} to the structural relaxation time $\tau_{\alpha}$
describing the decay of the intermediate scattering function.
Although the scaling of the two quantities is qualitatively similar,
the best fit exponent for $\tau_{\alpha}$ has the somewhat larger
value of $-2.8$.

The power law scaling of $S(\epsilon=1)$ with temperature is not
obvious since if the particle displacement were exactly described by
Brownian motion, then $\tau(\epsilon)$ would scale in inverse
proportion to the diffusion coefficient. A determination of the
diffusion coefficient for A particles is obtained by a simple
least-squares fit (not shown) of the long time data to the function
$\langle r^2 (t) \rangle = A + 6Dt$.  This fitting gives
\begin{equation} \label{dpowerlaw}
D \sim (T-T_c)^{2.1}.
\end{equation}
A more refined estimate by Kob and Anderson \cite{kob} on a smaller 
($1000$ particles) system
gave an exponent $2.0$ for the A particles.
The diffusion data thus scales with a fractional power of the structural 
relaxation time,
\begin{equation} \label{decoupling}
D \sim \tau_{\alpha}^{-(2.1/2.8)} \simeq \tau_{\alpha}^{-0.75}
\end{equation}
over the temperature range investigated.  Since evidence supports a
common temperature scaling of $\tau_{\alpha}$ and the fluid viscosity
$\eta$ \cite{onukiprl}, the observation implied by
Eq. (\ref{decoupling}) is consistent with the breakdown of the
Stokes-Einstein relation in real and simulated supercooled liquids
\cite{decoupling,footdec}. We therefore conclude that $S(\epsilon=1)$
scales neither exactly like the inverse structural relaxation time
$1/\tau_{\alpha}$ nor the diffusion coefficient $D$.  Other
characteristic times exist for this liquid. It has recently been
reported for the same simulations investigated here that the time
scale on which particle displacements are most correlated scales as a
power law with $(T-0.435)$, with an exponent $\gamma = 2.3 \pm 0.2$
\cite{dgp}. This exponent notably agrees within numerical error with
the exponent in Eq.~\ref{spowerlaw}.  We further note that the time
$t^*$ at which $\alpha(t)$ is a maximum (see Fig.~4) appears to
diverge at $T_c$ with an exponent $1.7$ \cite{dgpkp}.

\begin{figure}
\hbox to\hsize{\epsfxsize=1.0\hsize\hfil\epsfbox{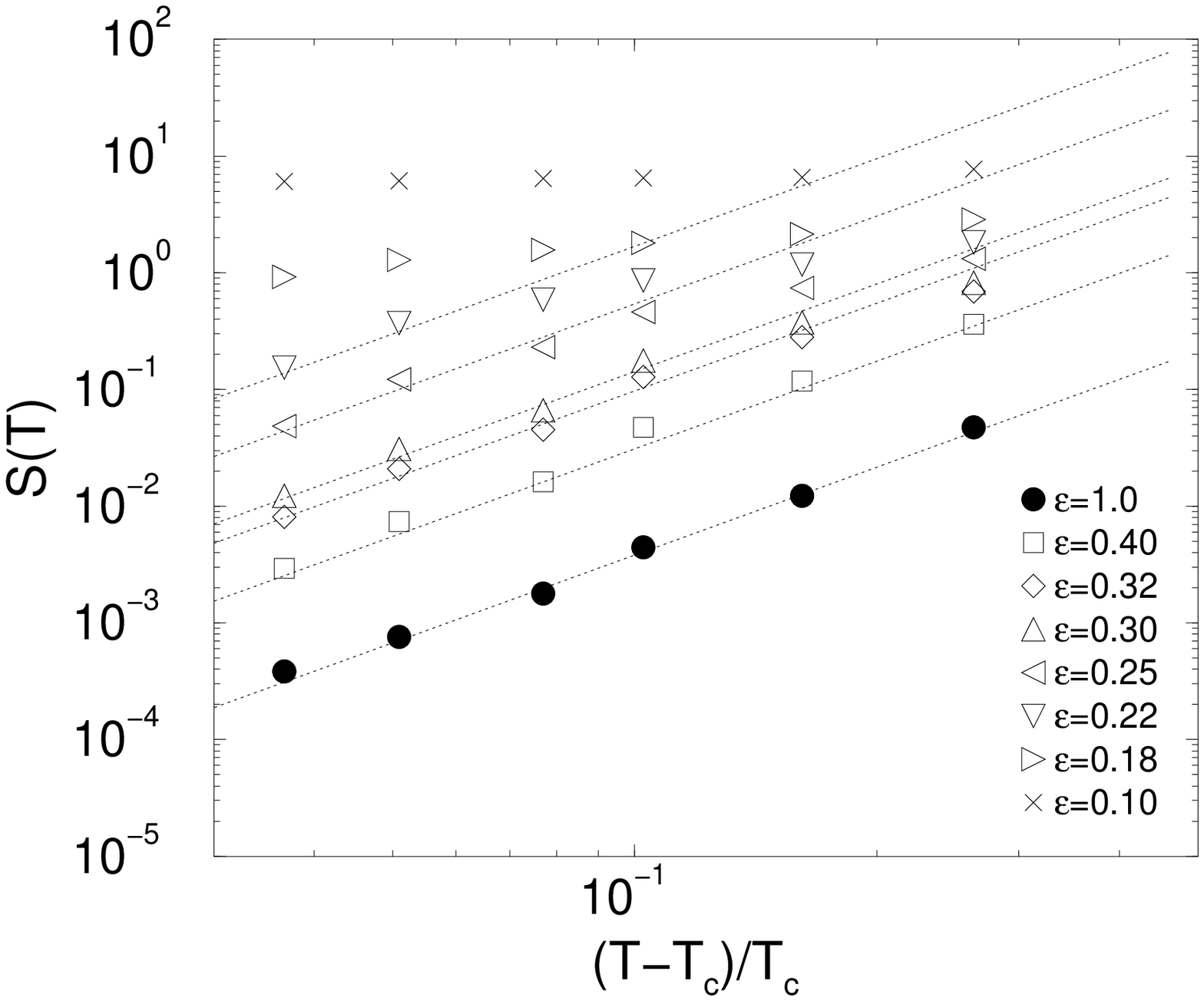}\hfil}
\caption{Mean first-passage time $\tau(\epsilon)$ versus $T-T_c$,
calculated at fixed values of $\epsilon$. Dotted lines denote the power law 
$S(T) \sim (T-T_c)^{2.5}$.}
\label{figvareps}
\end{figure}

Since the temperature dependence of $S(\epsilon)$ is largest in
Fig.~\ref{figtaueps} for $\epsilon=1$ and very small for $\epsilon
\rightarrow 0$, it is natural to consider the temperature dependence
of $S(\epsilon)$ for many fixed $\epsilon$ to determine how this
crossover occurs. In Fig.~\ref{figvareps} we see that the scaling $S(T) 
\propto (T-T_c)^{2.5}$ holds for $\epsilon$ greater than the cage size
$\epsilon_c$ and sufficiently small reduced temperature.  We also find
that for any temperature, the scaling fails for $\epsilon <
\epsilon_c$. This provides another method for determining the
cage size.

\section{Conclusion}
We have studied the length-scale dependence of the dynamic entropy
$S(\epsilon)$ in a molecular dynamics simulation of a model
supercooled binary Lennard-Jones liquid.  The simulations were
performed above both the glass transition temperature and the
mode-coupling critical temperature and correspond to equilibrated
liquid states \cite{dgpkp}.

$S(\epsilon)$ as estimated by the MFPT provides a tool for identifying
characteristic length and time scales of the dynamics of liquids and
provides a means for quantifying the degree of correlated motion
occurring in supercooled liquids. At very small $\epsilon$ we observe
a decorrelation of particle velocity and an $S(\epsilon)$ which is
insensitive to temperature and $\epsilon$ variation.  A decorrelation
time associated with an average interparticle collision time is
identified. At intermediate values of $\epsilon$ we observe a sharp
drop in $S(\epsilon)$ with increasing $\epsilon$, indicating that the
path motion is more ``stochastic'' at these length
scales. $S(\epsilon)$ is found to have a strong temperature dependence
as well in this regime. The logarithmic derivative $\Delta \equiv
d\log{\tau}/d\log{\epsilon}$ becomes a maximum at a characteristic
$\epsilon$ value that is identified with the particle ``cage'' size
$\epsilon_c$, since particle localization is maximal at this
point. The scaling of $S(\epsilon)$ with temperature at fixed
$\epsilon$ gives an independent confirmation of our estimation of
$\epsilon_c$ because it exhibits a qualitatively different dependence
on temperature for $\epsilon > \epsilon_c$ and $\epsilon <
\epsilon_c$.  The localization parameter $\Lambda \equiv
\Delta(\epsilon_c)$ increases and the cage size $\epsilon_c$ decreases
as the liquid is cooled.  The distribution functions for the
first-passage time at the scale of the cage size, $P_{\epsilon_c}(t)$,
in the coldest runs exhibit a long power-law tail, consistent with
suggestions that there is growing intermittency in the particle
displacements in glass-forming liquids \cite{odagaki,goodguy}. This
feature requires further study in cooler liquids to check the
predictions of these models.  At still larger scales (still less than
one interparticle separation) we observe persistent motion which
follows the transient particle ``caging''. $S(\epsilon)$ obeys a power
law $S(\epsilon) \sim \epsilon^{-1/\nu}$ for $\epsilon$ in the range
$(0.7,1)$ where the apparent fractal dimension of the particle
trajectories $\Delta(\epsilon) = 1/\nu$ ranges from $\approx 2$ to
$\approx 1.7$ as the temperature is lowered.  Thus, we observe a
tendency for the particle motion to acquire a persistent character
relative to Brownian motion as it is cooled.  This is consistent with
the previous observation of correlated string-like motion in this
liquid \cite{ddkppg}.

General arguments suggest that the dynamic entropy at microscopic
scales decreases at low temperatures, but a slower variation should be
obtained at higher temperatures, as in the present molecular dynamics
calculation (see Fig.~2). The situation is not so clear for the
first-passage time dynamic entropy $S(\epsilon)$ at the scale of one
interparticle separation $\epsilon=1$. In this case $S(\epsilon=1)$ is
inverse to the ``average interparticle exchange time \cite{frenkel}'',
$\tau(\epsilon=1)$.  We find that $S(\epsilon)$ vanishes as a power
law, $S(\epsilon=1) \propto (T-T_c)^{2.5}$ when the mode-coupling
temperature $T_c$ is approached from above. Thus, the cooled liquid
has the appearance of approaching an ergodic to non-ergodic transition
as $T \rightarrow T_c$, when viewed at the scale $\epsilon=1$. This is
consistent with the predictions of mode-coupling theory.

The first-passage time can also be utilized to obtain information
about the spatial dependence of mobility fluctuations in cooled
liquids. Perera and Harrowell \cite{perera}, for example, have
examined the position dependence of $\tau$ at the scale of one
interparticle spacing in a two-dimensional soft-sphere supercooled
liquid and found a tendency for particles of relatively high and low
``mobility'' (i.e., small and large $\tau$, respectively) to cluster
as $T$ is lowered. A detailed study of spatial correlations of
first-passage times in the present system will be presented elsewhere
\cite{ag}.

Future work should examine our approximate expression for
$S(\epsilon)$ in the $\epsilon \rightarrow 0$ limit through
independent calculation of the Kolmogorov-Sinai dynamic entropy,
$S(\epsilon \rightarrow 0) \equiv h_{KS}$. The temperature dependence
of $h_{KS}$ in cooler liquids ($T < T_c$) should be examined to
determine if there is a tendency for the ``bare'' dynamic entropy to
vanish at a lower glass transition temperature (see inset of
Fig.~12). Recent work has established a phenomenological relation
between dynamic entropy $h_{KS}$ and the equilibrium entropy in
simulations of model liquids at relatively elevated temperatures
\cite{dzugutov}. A decrease in $h_{KS}$ at lower $T$ should be
accompanied by the development of collective motion at short
times. Such motion has been reported in Ref.~\cite{hiwatari}. We
expect this change in the short time dynamics to be relevant for
interpreting the Boson peak phenomenon in cooled liquids
\cite{bosonpeak}. Simulations have already shown that the fraction of
unstable modes $f_u$ in a cooled liquid decreases in parallel with
$h_{KS}$ \cite{posch}, and the vanishing of $f_u$ has been identified
with the temperature where the diffusion coefficient $D$ vanishes
\cite{seeley,sciortino}. 

Finally, we note that the calculation for $S(\epsilon)$ can be
extended to other dynamical variables associated with other transport
properties (viscosity, thermal conductivity, etc.) and these
calculations should provide useful estimates of other characteristic
space and time scale in cooled liquids \cite{DorfmanGaspard}. We
emphasize that although the definition of $S(\epsilon)$ is motivated
by dynamical systems theory concepts, this quantity defines an
independently interesting measure of correlated motion in liquids that
does not rely on the approximation relating $S(\epsilon)$ to
$h(\epsilon)$.

\bigskip
\noindent {\it Corresponding author:} {\bf sharon.glotzer@nist.gov}.


\begin{thebibliography}{99}
\bibitem{sametc} Recall that for this same system simulated along a
different path, the same value of $T_c$ was found from an analysis of
the minority (B) particles only \cite{kob}.
\end{thebibliography}
\end{document}